\documentclass[aps,prx,twocolumn,superscriptaddress]{revtex4-1}

\usepackage{epsfig}
\usepackage{amssymb} 
\usepackage{amsfonts} 
\usepackage{mathtools} 
\usepackage{dcolumn} 
\usepackage{graphicx} 
\usepackage{color}  
\usepackage[normalem]{ulem}
\usepackage{bm}  
\usepackage{comment} 
\usepackage[switch]{lineno}
\usepackage{units}
\usepackage{sidecap} 
\usepackage{textcomp} 
\usepackage{mathrsfs} 

\definecolor{darkblue}{rgb}{0.0,0.0,0.6}

\definecolor{darkgreen}{rgb}{0.02,0.45,0.0}

\definecolor{violet}{rgb}{0.8,0.2,0.6}

\usepackage{array} 
\usepackage{upgreek}
\usepackage{tabularx}
\usepackage{latexsym}
\usepackage{bm,array}
\usepackage{amsfonts}
\usepackage{amssymb}
\usepackage{amsmath}
\usepackage{bigstrut}
\usepackage{placeins}
\usepackage[usenames,dvipsnames]{xcolor}
\usepackage[bookmarks=false,colorlinks]{hyperref}
\hypersetup{
	linkcolor=MidnightBlue,        
	citecolor=blue,     
	filecolor=Plum,      	
	urlcolor=MidnightBlue,           
}

\newcommand{\bpm}{\begin{pmatrix}}
\newcommand{\epm}{\end{pmatrix}}
\newcommand{\be}{\begin{eqnarray}}
\newcommand{\ee}{\end{eqnarray}}
\newcommand{\ba}{\begin{array}}
\newcommand{\ea}{\end{array}}

\def \nio{Na$_2$IrO$_3$}
\def \anio{$\alpha$-Na$_2$IrO$_3$}
\def \alio{$\alpha$-Li$_2$IrO$_3$}
\def \blio{$\beta$-Li$_2$IrO$_3$}
\def \glio{$\gamma$-Li$_2$IrO$_3$}

\begin{document}

\title{Magnon-spinon dichotomy in the Kitaev hyperhoneycomb $\beta$-Li$_2$IrO$_3$}

\author{Alejandro Ruiz}
\email[Corresponding Author:]{alejandro@ucsd.edu}
\affiliation{Department of Physics, University of California, San Diego, California 92093, USA}

\author{Nicholas P. Breznay}
\affiliation{Department of Physics, Harvey Mudd College, Claremont, CA 91711}

\author{Mengqun Li}
\affiliation{School of Physics and Astronomy, University of Minnesota, Minneapolis, Minnesota 55116, USA}

\author{Ioannis Rousochatzakis}
\affiliation{Department of Physics, Loughborough University, Loughborough, LE11 3TU, United Kingdom }

\author{Anthony Allen}
\affiliation{Department of Physics, University of California, San Diego, California 92093, USA}

\author{Isaac Zinda}
\affiliation{Department of Physics, Harvey Mudd College, Claremont, CA 91711}

\author{Vikram Nagarajan}
\affiliation{Department of Physics, University of California, Berkeley, California 94720, USA}
\affiliation{Materials Sciences Division, Lawrence Berkeley National Laboratory, Berkeley, California 94720, USA}

\author{Gilbert Lopez}
\affiliation{Department of Physics, University of California, Berkeley, California 94720, USA}
\affiliation{Materials Sciences Division, Lawrence Berkeley National Laboratory, Berkeley, California 94720, USA}

\author{Mary H. Upton}
\affiliation{Advanced Photon Source, Argonne National Laboratory, Argonne, Illinois 60439, USA}

\author{Jungho Kim}
\affiliation{Advanced Photon Source, Argonne National Laboratory, Argonne, Illinois 60439, USA}

\author{Ayman H. Said}
\affiliation{Advanced Photon Source, Argonne National Laboratory, Argonne, Illinois 60439, USA}

\author{Xian-Rong Huang}
\affiliation{Advanced Photon Source, Argonne National Laboratory, Argonne, Illinois 60439, USA}

\author{Thomas Gog}
\affiliation{Advanced Photon Source, Argonne National Laboratory, Argonne, Illinois 60439, USA}

\author{Diego Casa}
\affiliation{Advanced Photon Source, Argonne National Laboratory, Argonne, Illinois 60439, USA}

\author{Robert J. Birgeneau}
\affiliation{Department of Physics, University of California, Berkeley, California 94720, USA}
\affiliation{Materials Sciences Division, Lawrence Berkeley National Laboratory, Berkeley, California 94720, USA}

\author{Jake D. Koralek}
\affiliation{SLAC National Accelerator Laboratory, Menlo Park,  California, 94025, USA}

\author{James G. Analytis}
\affiliation{Department of Physics, University of California, Berkeley, California 94720, USA}
\affiliation{Materials Sciences Division, Lawrence Berkeley National Laboratory, Berkeley, California 94720, USA}

\author{Natalia B. Perkins}
\affiliation{School of Physics and Astronomy, University of Minnesota, Minneapolis, Minnesota 55116, USA}

\author{Alex Frano}
\email[Corresponding Author:]{afrano@ucsd.edu}
\affiliation{Department of Physics, University of California, San Diego, California 92093, USA}

\date{\today}

\begin{abstract}
The family of edge-sharing tri-coordinated iridates and ruthenates has emerged in recent years as a major platform for Kitaev spin liquid physics, where spins fractionalize into emergent magnetic fluxes and Majorana fermions with Dirac-like dispersions.
While such exotic states are usually pre-empted by long-range magnetic order at low temperatures, signatures of Majorana fermions with long coherent times have been predicted to manifest at intermediate and higher energy scales, similar to the observation of spinons in quasi-1D spin chains. 
Here we present a Resonant Inelastic X-ray Scattering study of the magnetic excitations of the hyperhoneycomb iridate \blio{} under a magnetic field with a record-high-resolution spectrometer. At low-temperatures, dispersing spin waves can be resolved around the predicted intertwined incommensurate spiral and field-induced zigzag orders, whose excitation energy reaches a maximum of \unit[16]{meV}. A 2\,T magnetic field softens the dispersion around ${\bf Q}=0$. The behavior of the spin waves under magnetic field is consistent with our semiclassical calculations for the ground state and the dynamical spin structure factor, which further predicts that the ensued intertwined uniform states remain robust up to very high fields (100\,T). Most saliently, the low-energy magnon-like mode is superimposed by a broad continuum of excitations, centered around \unit[35]{meV} and extending up to \unit[100]{meV}. This high-energy continuum survives up to at least \unit[300]{K} -- well above the ordering temperature of \unit[38]{K} -- and gives evidence for pairs of long-lived Majorana fermions of the proximate Kitaev spin liquid. 
\end{abstract}

\pacs{71.18.+y,74.72.-h,72.15.Gd}
\maketitle
\section{Introduction}
Recent years have seen a vigorous experimental research effort on identifying candidate materials for the celebrated quantum spin liquid (QSL)~\cite{anderson_resonating_1987,Wen2002,Balents2010,Savary2016,Takagi2019,Broholm2020,Motome2020a}, a topological state of matter with long-range entanglement and fractionalized excitations, whose efficient control can lead to novel and transformative technological applications~\cite{Kitaev2003,kitaev_anyons_2006,Das2008,Aasen2020}.  
A central focus in this effort are the tri-coordinated Kitaev materials~\cite{jackeli_mott_2009,BookCao,rau_review_2016,Trebst2017, Winter2017,Knolle2017,Takagi2019,Motome2020a}, a family of spin-orbit-assisted Mott insulators, in which spin-orbit-entangled $j_\mathrm{eff}\!=\!1/2$ magnetic moments interact via bond-directional Ising-like interactions, known as Kitaev interactions~\cite{kitaev_anyons_2006}. 
When acting alone, these interactions lead to a variety of exactly solvable QSLs, whose elementary excitations -- Majorana fermions moving on the background of emergent magnetic fluxes~ \cite{kitaev_anyons_2006, Mandal2009,Hermanns2015} -- give rise to characteristic broad signatures in dynamical response functions~\cite{Knolle2014a,knolle_raman_2014,Knolle2015,Gabor2016,Gabor2017,halasz_fract_2019,Rousochatzakis_finite_2019}.

Most candidate materials, however,  exhibit ordered states at low temperatures instead of the sought-after QSL. Known examples include
the collinear, commensurate zigzag state  in Na$_2$IrO$_3$~\cite{Hill2011,choi_spin_2012} and $\alpha$-RuCl$_3$~\cite{johnson_rucl3_2015}, and the non-coplanar, counter-rotating incommensurate spiral order in $\alpha,\beta$,\glio~\cite{biffin_noncoplanar_2014,biffin_unconventional_2014, williams_incommensurate_2016}.
Such ordered states arise due to the presence of additional interactions, such as the Heisenberg exchange $J$ and the symmetric off-diagonal anisotropy $\Gamma$~\cite{jackeli_mott_2009,chaloupka_kitaev-heisenberg_2010,Katukuri2014,Rau2014,Sizyuk2014,lee_theory_2015,lee_3d_2016,Winter2017}.
While such perturbations naturally impede the realization of QSLs, a wealth of theoretical and experimental work has shown that the Kitaev exchange is still the dominant interaction in all Kitaev materials, and that it strongly influences their magnetic behavior \cite{hwan_chun_direct_2015,rau_review_2016,Winter2017}. 
Most saliently, it has been suggested that this separation of energy scales must give rise to an extended {\it `proximate spin-liquid' regime} with signatures of long-lived fractionalized excitations, separating the low-$T$ ordered phase from the high-$T$ paramagnetic regime~\cite{nasu_vaporization_2014,nasu_thermal_frac_2015,nasu_fermionic_2016, nasu_thermal_2017,banerjee_excitations_2018,kasahara_majorana_2018,halasz_fract_2019,Rousochatzakis_finite_2019}.

For example, inelastic neutron scattering measurements in RuCl$_3$ have revealed a gapped, broad continuum of magnetic excitations far above the magnetic ordering temperature~\cite{banerjee_proximate_2016,banerjee_neutron_2017}, while Raman spectroscopy has provided evidence for the fermionic character of these excitations~\cite{sandilands_scattering_2015,glamazda_raman_2016,wang_range_2020,Li_LIO_ramam_2020}. 
These observations signify a dichotomy between the expected magnon-like response at low energies, originating from the long-range ordering, and a continuum multi-spinon-like response at higher energies, originating from pairwise excitations of long-lived Majorana fermions of the proximate Kitaev QSL model. 
This proximate spin-liquid regime 
is characterized by strong, short-range spin-spin correlations and disordered fluxes, and can be observed in thermodynamic measurements, or more directly in dynamical probes such as inelastic neutron scattering, Raman and ultrafast spectroscopy, and Resonant Inelastic x-ray Scattering (RIXS)~\cite{nasu_vaporization_2014,nasu_thermal_frac_2015,nasu_fermionic_2016, nasu_thermal_2017,banerjee_excitations_2018,kasahara_majorana_2018,halasz_fract_2019,Rousochatzakis_finite_2019}.

\begin{figure*}[ht]
{\includegraphics[width=\textwidth]{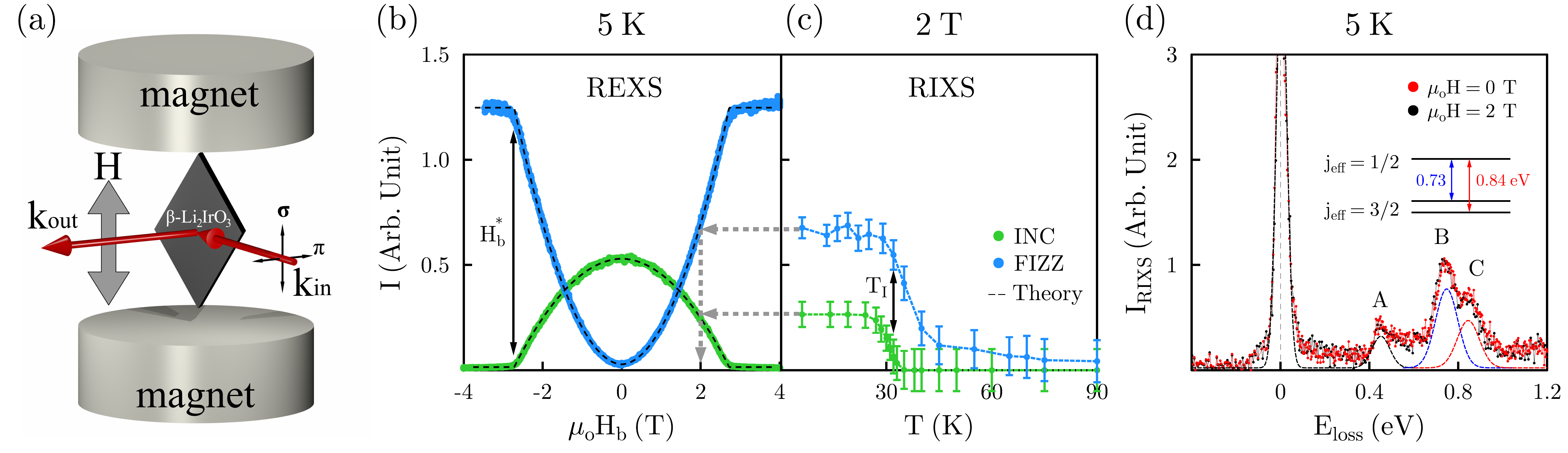}
\caption{(Color online) (a) The \blio{} single crystal oriented with $\hat{a}-$axis in the scattering plane and the $\hat{b}-$axis perpendicular to it. The crystal was placed between two neodymium permanent magnets, with a \unit[2]{T} magnetic field applied along the $\hat{b}-$axis and a $\pi$-polarized x-ray beam. 
(b) The field dependence of the scattered intensity at \unit[5]{K} measured at the wave vectors corresponding to IC and FIZZ (${\bf Q}=(0.574,0,22)$ and ${\bf Q}=(0,0,22)$) states using a DC magnetic field~\cite{ruiz_correlated_2017}, to be compared with the intensity measured in the RIXS setup--shown in panel (c)--for calibration of the field value. The dotted lines represented the calculated Bragg peak intensities~ \cite{rousochatzakis_magnetic_2018}. 
(c) The temperature dependence (onset temperature) of the integrated RIXS intensity of the intertwined orders confirms the magnitude of the applied field ($\mu_o$H$_b$ $\approx\unit[2]{T}$). 
(d) Wide energy range RIXS spectra at ${\bf Q}=(0.2,0,22)$ obtained with incident energy $E_i=\unit[11.2145]{keV}$ for two magnetic field values. Feature A ($\approx\unit[0.42]{eV}$) corresponds to an intersite exciton formed by a particle-hole pair across the Mott gap, while B ($\approx\unit[0.73]{eV}$) and C ($\approx\unit[0.84]{eV}$) represent intrasite excitations of $3\lambda_{SO}/2$ energy between $j_\mathrm{eff}\!=\!\nicefrac{1}{2}$ and $j_\mathrm{eff}\!=\!\nicefrac{3}{2}$ states, indicating a spin-orbit coupling $\sim\unit[0.5]{eV}$. No measurable change in the energy-loss spectrum is observed as a result of the applied $\unit[2]{T}$ field. }
\label{fig:fig1}} 
\end{figure*}

Here we report experimental RIXS evidence for such a dichotomy of low-energy magnons and broad excitations from the proximate Kitaev spin liquid regime in the 3D hyperhoneycomb iridate \blio{}, with a characteristic continuum response extending up to at least 300 K, well above the ordering temperature of 38 K in this material.

\blio{} is perhaps the most intriguing example of the complex interplay between the Kitaev exchange and external magnetic fields \cite{ruiz_correlated_2017,majumder_nmr_2019, Majumder2020}. At zero field, the system orders at $T_I\!=\!\unit[38]{K}$ into a complex incommensurate order (IC) with counter-rotating spirals and propagation wavevector ${\bf Q}\!=\!(0.574,0,0)$; all ${\bf Q}-$vectors are reported in orthorhombic reciprocal lattice units~\cite{biffin_unconventional_2014}. Applying a magnetic field along the crystallographic $\hat{b}$-axis rapidly destroys the IC order at a critical field $\mu_o$H$^*_b\!=\!\unit[2.8]{T}$, where the system is driven into a uniform quantum correlated coplanar phase, comprising a ferromagnetic (FM) component along the field and a zigzag component along $\hat{a}$, similar to the magnetic order of \nio\, and RuCl$_3$ \cite{ruiz_correlated_2017}.  In the following, we refer to this phase as a field-induced zigzag (FIZZ) phase.

It is now understood~\cite{ducatman_2018,rousochatzakis_magnetic_2018,Li_magneticfield_2020,li_reentrant_2020} that the IC order of $\beta$-Li$_2$IrO$_3$ can be thought of as a long-wavelength twisting of the energetically closest commensurate state with ${\bf Q}\!=\!(\frac{2}{3},0,0)$, which is strongly intertwined with a set of uniform orders, both at zero and non-zero magnetic fields.  
In addition, most of the available experimental data can be described by a minimal microscopic $J$-$K$-$\Gamma$ model~\cite{lee_theory_2015,lee_3d_2016}, with coupling parameters $J\!\simeq\!0.4$~meV, $K\!\simeq\!-18$~meV, and $\Gamma\!\simeq\!-10$~meV
 \cite{Li_magneticfield_2020,li_reentrant_2020}. 
Despite this understanding, there is very little experimental information about the low-energy excitation spectra and their response to magnetic fields, in part because the crystal size and the large neutron absorption cross-section of Ir impede inelastic neutron scattering studies.

The experiments reported below provide a comprehensive picture of the dynamic response of \blio{} under magnetic fields H$_b$ along the $\hat{b}$-axis. We measured its low-energy magnetic excitation spectrum by performing Ir-$L_3$-edge RIXS experiments. Central to this effort are complementary experiments with a conventional spectrometer with a medium energy resolution (MER) $\Delta E\sim\unit[25]{meV}$ as well as a state-of-the-art spectrometer with a high energy resolution (HER) $\Delta E\sim\unit[10]{meV}$. These measurements were performed at \unit[0]{T} as well as under a \unit[2]{T} magnetic field. The results of theses measurements show magnetic excitations branching from the magnetic zone centers corresponding to the intertwined IC and FIZZ states. We  observe that magnetic excitations naturally split into low-energy dispersive modes and high-energy continuum like excitations. The former reach a maximum energy around \unit[16]{meV}, and an applied 2\,T magnetic field softens the dispersion around ${\bf Q}=0$.
The experimental data are in very good agreement with semiclassical calculations for the dynamical spin structural factor of \blio{}, lending strong support to the microscopic $J$-$K$-$\Gamma$ description. 
%
Furthermore, we identify a broad continuum of magnetic excitations whose intensity is insensitive to the low-temperature magnons, remains constant up to $\sim\unit[100]{K}$ 
(i.e., well above $T_I=\unit[38]{K}$), and slowly decreases at higher temperatures. The temperature dependence of this continuum is consistent with the behavior of nearest neighbor spin correlations emerging from the dominant Kitaev energy scale, and its persistence to high temperature alludes to the long coherence times of the fractionalized excitations of the proximate QSL phase~\cite{Gabor2016,Gabor2017,halasz_fract_2019}.

The paper is organized as follows: Section II describes the crystal synthesis and the RIXS experiment set-up.  Section III presents the experimental data and calculations of the spin dynamical structural factor. Section IV discusses the strongly interwined nature of the IC and FIZZ states, and the origin of the multi-spinon continuum. Section V summarizes our findings. 

\section{Experimental Details}
High-quality single crystals of \blio\ were grown by a vapor transport technique. Ir (99.9\% purity, BASF) and Li$_2$CO$_3$ (99.999 \% purity, Alfa-Aesar) powders were ground and pelletized at \unit[3,000]{psi} in the molar ratio of 1:1.05. The pellets were placed in an alumina crucible, reacted for 12 h at 1,050\textdegree C, and then cooled to room temperature at 2 \textdegree C/h to yield single crystals which were then extracted from the reacted powder. \blio{} crystallizes in the orthorhombic Fddd space group with single-crystal sizes $\sim$ 100$\times$150$\times$300 $\upmu$m$^3$ (more details in the Supplementary Information).

RIXS measurements were performed at beamline 27-ID of the Advanced Photon Source using $\pi$-polarized x-rays in the horizontal scattering plane at detector angles close to 2$\theta\sim90^{\circ}$, in order to suppress elastic scattering. The RIXS spectra were obtained using two different setups: 
(1) a medium-resolution $\Delta E_{\textrm{MER}}\sim\unit[25]{meV}$ with a two-bounce Si(844) monochromator and a 2m-radius Si(844) diced spherical analyzer \cite{GogJoSR2013}, and (2) a high-resolution $\Delta E_{\textrm{HER}}\sim\unit[10]{meV}$ with a four-bounce Si(844) monochromator and a 2m-radium quartz(309) diced spherical analyzer \cite{Said:hf5355,gog_performance_2018} at the $Ir$ L$_3$ absorption edge (2$p_{3/2}\rightarrow5d_{t_{2g}}\sim \unit[11.215]{keV}$). 


To study the magnetic field dependence of the low-energy excitations, we devised the setup shown in Figure \ref{fig:fig1}\,a. A \blio\, crystal with a surface normal in the [001] crystallographic direction was mounted between two Nd$_2$Fe$_{14}$B magnets separated by 400~$\mu$m. The scattering plane is defined by the (001)$\times$(100) reciprocal lattice vectors, and the magnetic field is applied parallel to the [010] direction ($\hat{b}$-axis). 

\begin{figure*}[ht]
    {\includegraphics[width=\textwidth ]{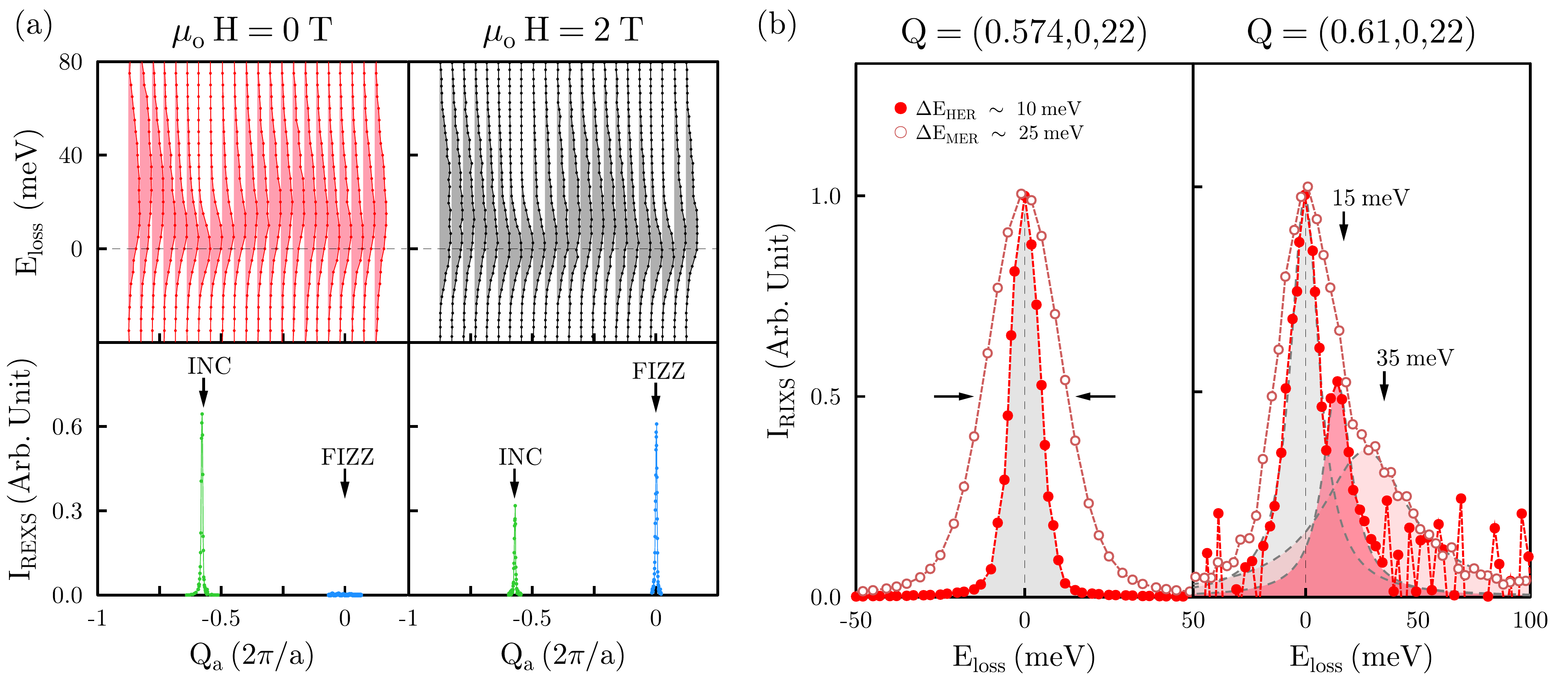}
  	\caption{(Color online) (a) Upper panels show the momentum dependence of the RIXS spectra taken along the  (h,0,22) direction at \unit[5]{K} with \unit[0]{T} and \unit[2]{T} fields. Lower panels show the REXS scans around the zone centers of the INC and FIZZ states. The applied magnetic field suppresses the INC order while giving rise to FIZZ state \cite{ruiz_correlated_2017}. (b) A comparison of the low-energy RIXS spectra collected at \unit[5]{K} with two spectrometers. In the vicinity of the INC Bragg peak ${\bf Q}=(0.574, 0, 22)$, the FWHM of the elastic peak is $\Delta E_{\textrm{MER}}=\unit[25]{meV}$ and $\Delta E_{\textrm{HER}}=\unit[10]{meV}$, respectively. Slightly away from the Bragg peak at ${\bf Q}=(0.61, 0, 22)$, two inelastic modes are identified: a dispersing magnon with a maximum energy of $\sim\unit[16]{meV}$, and a broad continuum of magnetic excitations centered $\sim\unit[35]{meV}$ which is nearly momentum-independent.}
  	\label{fig:fig2}} 
\end{figure*}

\section{Results}

Measurements of the resonant elastic x-ray scattering (REXS) intensity taken with a tunable DC magnetic field \cite{ruiz_correlated_2017} on the same sample serve as a check of the magnetic field experienced by the sample in the permanent magnet set-up. Figure \ref{fig:fig1}\,b,c show the comparison of the low-temperature REXS intensity with the RIXS integrated intensity taken at the same reciprocal space points corresponding to the IC (${\bf Q}=(0.574, 0, 22)$) and the FIZZ ${\bf Q}=(0, 0, 22)$ states.
 
 The magnetic field suppresses the non-coplanar IC state that develops below $T_I=\unit[38]{K}$ \cite{biffin_noncoplanar_2014}, while the system develops a coplanar FIZZ component~\cite{ruiz_correlated_2017}. 
The calculated field evolution of the Bragg peak intensities are also shown in Figure \ref{fig:fig1}\,b. The low temperature integrated intensity values of both peaks measured in our RIXS setup map directly onto the REXS intensity values for a $\unit[2]{T}$ field, shown in Figure \ref{fig:fig1}\,c. Furthermore, the temperature dependence of the integrated RIXS signal is plotted on Figure \ref{fig:fig1}\,c. The observed transition temperature in the in-field RIXS data, $T_{I}(\textrm{H})=\unit[32]{K}$, is consistent with an applied $\unit[2]{T}$ field when compared to the reported thermodynamic measurements \cite{takayama_hyperhoneycomb_2015,ruiz_correlated_2017, majumder_nmr_2019}.

\begin{figure*}[ht]
{\includegraphics[width=\textwidth]{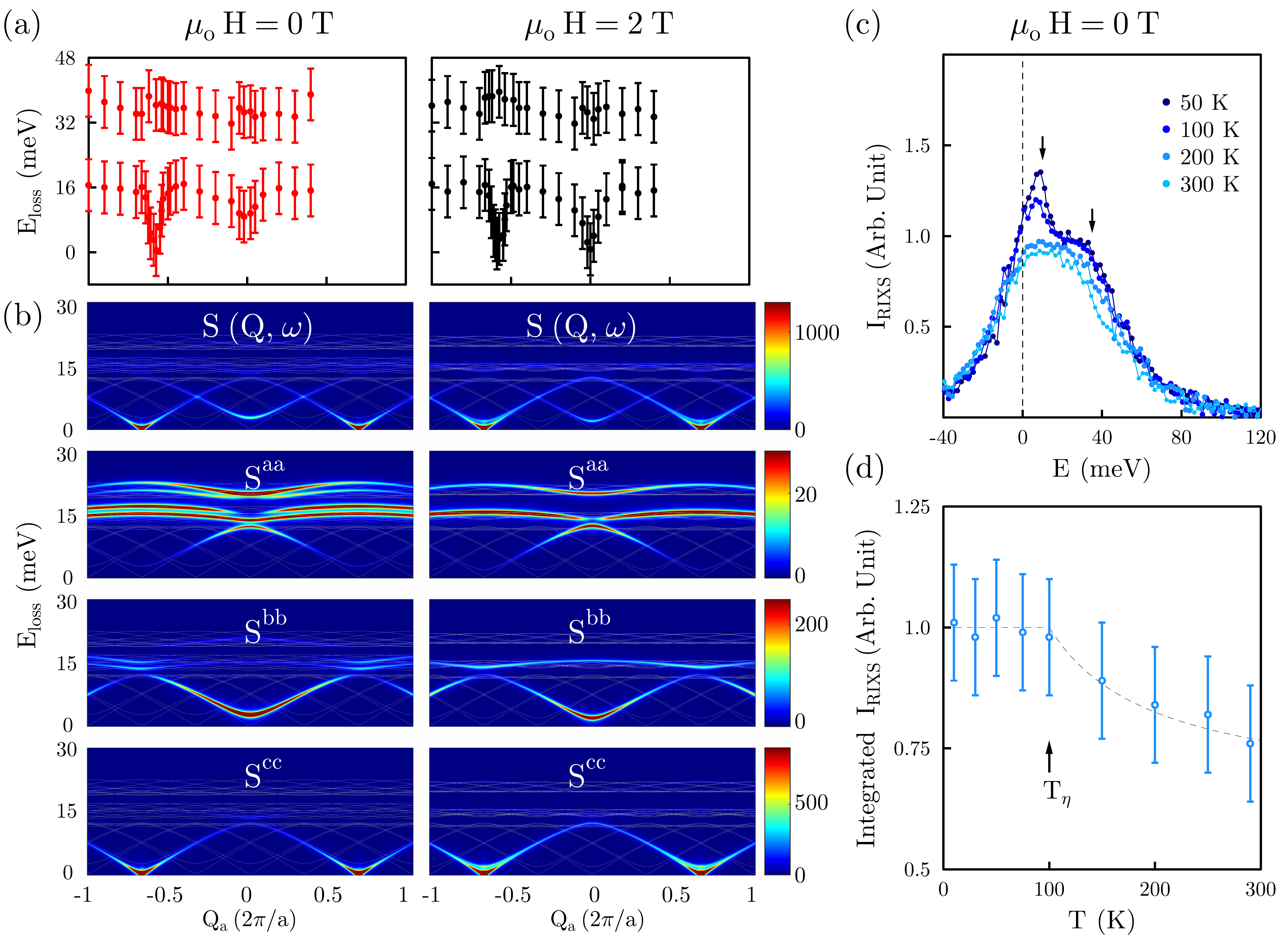}
\caption{(Color online) 
(a) The momentum dependence of the low-energy RIXS excitations was extracted from the data in Figure \ref{fig:fig2}\,a. The magnon mode with a maximum energy $\sim\unit[16]{meV}$ disperses toward the zone centers corresponding to the IC ($Q_a=0.574$) and FIZZ ($Q_a=0$) states. The \unit[2]{T} spectra softens around $Q_a=0$ compared to the \unit[0]{T} data. The experimental results are well captured by the dynamical structure factor calculations presented in panel (b). On the other hand, the broad magnetic continuum centered around \unit[35]{meV} is insensitive to momentum transfer along $Q_a$. 
(b) The diagonal components of the spin dynamical structure factor, $S^{aa}(Q,\omega),\, S^{bb}(Q,\omega),\,  S^{cc}(Q,\omega)$, and their sum, $S (Q,\omega)$, computed  at H$_{\bf b}=0\,$T (left panels) and H$_{\bf b}=2\,$T (right panels) along the $Q_a$ direction. The colors and the width indicate the magnitude of the associated component after convolving with a Gaussian. Note that the color scale  shown individually for each panel  varies significantly between different components. 
(c) Exemplary temperature dependence of the 0\,T RIXS excitation spectra above $T_I$. (d) The 0\,T RIXS intensity integrated above $\unit[20]{meV}$ shows that the inelastic continuum hardly changes up to \unit[100]{K} and persists up to \unit[300]{K}. This feature seems to be insensitive to the low-temperature INC transition at $T_I$.}
  	\label{fig:fig3}} 
\end{figure*}

Figure \ref{fig:fig1}\,d shows the MER RIXS spectra as a function of energy loss ($E_{loss}=E_i-E_f$) up to \unit[1.2]{eV}. These scans were performed at \unit[5]{K} near ${\bf Q} = (0.2,0,22)$, a location well away from any Bragg reflection. We observe three features, labeled  A, B and C, which were fitted using three Gaussian functions on top of a broad Gaussian background. Feature A corresponds to intersite particle-hole excitations of the $j_\mathrm{eff}\!=\!\nicefrac{1}{2}$ band across the Mott gap. Features B and C correspond to intrasite excitations between the $j_\mathrm{eff}\!=\!\nicefrac{1}{2}$ and $j_\mathrm{eff}\!=\!\nicefrac{3}{2}$ states~ \cite{Rotenberg2008,Gretarson_CFS_2012}, resulting from a spin-orbit coupling $\sim\unit[0.5]{eV}$. A small difference between B and C energies indicates the presence of the trigonal crystal field splitting.
The applied \unit[2]{T} magnetic field has no noticeable effect on the energy nor intensity of these electronic excitations; hence the localized $j_\mathrm{eff}=\nicefrac{1}{2}$ pseudo-spin provides a suitable description for the electronic ground state of \blio\, under an applied magnetic field. This is in stark contrast with recent observations of the collapse of the relativistic $j_\mathrm{eff}=\nicefrac{1}{2}$ state under the application of remarkably small hydrostatic pressure \cite{clancy_pressure-driven_2018,takayama_pressure_2019}.

\begin{figure*}[ht]
    {\includegraphics[width=\textwidth]{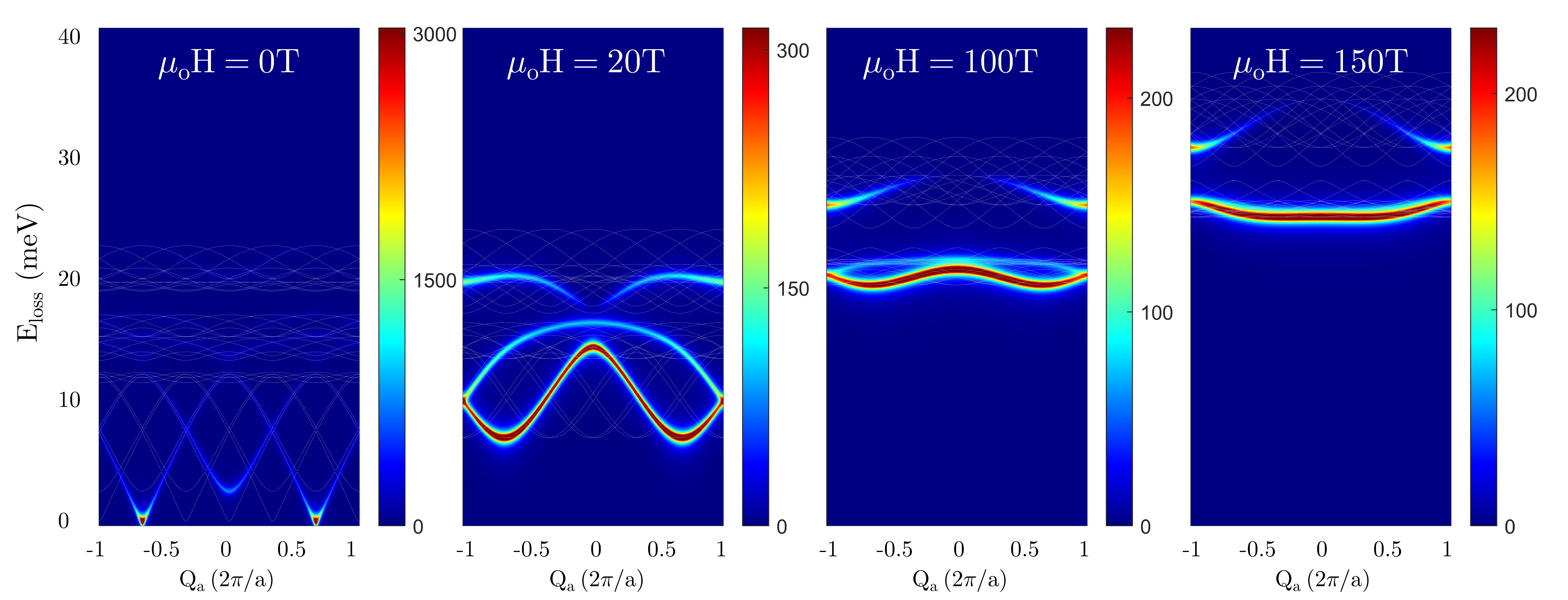}
  	\caption{(Color online) Evolution of the total dynamical structure factor $\mathcal{S}({\bf Q},\omega)$ calculated in magnetic field along the ${\bf b}$ axis. The spectra is shown for $H_{\bf b}=0\text{T},\,20\text{T},\,100\text{T},$ and $150\text{T}$. Even for $H_b\gg H_b^*$, the $Q_a=2/3$ mode remains soft and significantly intense, indicating that IC and FIZZ states are intertwined at all fields.  The colors and the width indicate the magnitude of the associated component after convolving with a Gaussian.}
  	\label{fig:fig4}} 
\end{figure*}

Having established that the field does not alter the spin-orbit entangled Mott state, we investigate its effect on the low-energy excitations. The top panel in Figure \ref{fig:fig2}\,a compares raw MER spectra for 0 and 2\,T over a limited energy loss window. These data were taken at different momentum transfer positions with varying in-plane value ${\bf Q_a}=(h,0,22)$. The spectra are normalized to their respective intensity maxima for better visualization. A clear dispersion can be observed as the inelastic features appear to soften around the respective IC and FIZZ ${\bf Q}$-vectors. The bottom panels show the integrated REXS signal for both field values~\cite{ruiz_correlated_2017}, revealing sharp Bragg peaks that span a much narrower range than that over which the inelastic features soften. Intriguingly, the \unit[2]{T} spectrum near ${\bf Q}=(0,0,22)$ has shifted to lower energy compared to the \unit[0]{T} data, which constitutes one of the central findings of our work.

To better understand these low-energy features, Figure \ref{fig:fig2}\,b shows representative low-energy RIXS spectra for \blio{}, collected at \unit[5]{K} using the MER and HER spectrometers. The first panel shows the spectra taken at the IC Bragg peak position and was used to determine the working resolution. The strong elastic signal has a full-width-at-half-maximum (FWHM) of \unit[25]{meV} and \unit[10]{meV} respectively, consistent with previous calibration measurements \cite{gog_performance_2018, kim_rixs_2020}. The second panel compares inelastic spectra taken at ${\bf Q}=(0.61,0,22)$, a reciprocal lattice position away from the IC ${\bf Q}$-vector. In the MER scans, a clear shoulder of intensity is seen at non-zero energy loss. These spectra were fitted by a sum of pseudo-Voigt functions representing the elastic line (gray) and two inelastic features (red and pink) $\sim\unit[15]{meV}$ and $\sim\unit[35]{meV}$. The elastic line width was set by the energy resolution and only its intensity and peak position was allowed to vary in the fit (more details in the Supplementary Information). Since the lowest energy inelastic feature (red) is within the energy resolution of the MER scan, we compare them to HER scans taken at the same reciprocal space positions. The inelastic signal $\sim$ 15\,meV  is most clearly seen in the high resolution spectra. (The mode at $\sim$35\,meV is not visible in the HER scans because of the lower throughput of this detector.) In summary, the data show (i) a resolution-limited, low-energy excitation ($E_{loss}<\unit[16]{meV}$) which is sensitive to momentum transfer, and (ii) a broad continuum centered around \unit[35]{meV} with width much larger than the experimental resolution and insensitivity to variations along the $Q_a$ momentum transfer direction. This dichotomy is the key experimental finding of our work. 

These low-energy inelastic features can originate from either lattice or spin excitations within the $\sim\unit[400]{meV}$ Mott gap \cite{hinton_lio_2014}. Phonon excitations arise due to dynamics in the short-lived intermediate state and are expected to be very weak in the well-screened intermediate state of Ir L-edge RIXS, which explains why no detectable phonon contributions have thus far been observed in other Mott insulator $5d^5$ iridates \cite{RIXS_2011,revelli_fingerprints_2020}. Moreover, O K-edge RIXS experiments on Li$_2$IrO$_3$ showed that the majority of the phonon spectra resides around 70 meV \cite{Vale_OK_RIXS_2019}, well above the inelastic features we observed. 

By contrast, magnetic excitations are strongly enhanced by the resonance process. In particular, the magnetic contribution to the RIXS response comes from two distinct types of processes, the spin-conserving (SC) and the non-spin-conserving (NSC)~\cite{Gabor2016,Gabor2017}. In a SC process, the spin of the 5d valence shell does not change during RIXS, while in a NSC process the spin gets flipped. Furthermore, for the pure Kitaev model, there are three types of NSC channels, corresponding to the rotation of the spin by $\pi$ around the $x$, $y$, and $z$ axes, respectively. These three NSC channels and the SC channel are orthogonal in the pure Kitaev spin liquid  \cite{Gabor2016,Gabor2017}, but 
they are not fully orthogonal in the more complete, $J$-$K$-$\Gamma$ description of \blio{}. 
Nevertheless,  we can assume that  since the Kitaev interaction is dominant, part of the response is predominantly coming from the NSC channel and part from the SC channel. In this case, the NSC RIXS response reduces to the corresponding diagonal component of the dynamical spin structure factor (DSF)  defined as
\begin{align}\label{DSF-main}
\mathcal{S}^{\alpha\beta}({\bf Q},\omega)&=\!\!\int\!\! \textit{}{dt} \, e^{-i\omega t}\langle S^\alpha(-{\bf Q},0)S^\beta({\bf Q},t)\rangle,
\end{align}
where ${\bf Q}$ and $\hbar \omega$ are the momentum and energy transfers, and 
\begin{align}
{ S}^\alpha({\bf Q},t) =\frac{1}{\mathcal{N}} \sum_{{\bf r}} e^{-i {\bf Q}\cdot{\bf r}} S^\alpha_{\bf r}(t)
\end{align}
is the Fourier transform of the spin density at time $t$, with $\mathcal{N}$ being the total number of spins and ${\bf r}$ denoting the physical positions of the spins. 
The SC RIXS channel will manifest itself only if fractionalized excitations are present. It will pick up exclusively the Majorana fermions of the fractionalized Kitaev spin liquid, and the overall energy dependence of this contribution will be proportional to the two-fermion joint density of states \cite{Gabor2016,Gabor2017} (more details in the Supplementary Information).

Let us return to our RIXS data.
To probe the nature of the magnetic contribution to the RIXS response of \blio{}, we extract the full momentum and temperature dependence. Figure \ref{fig:fig3}\,a shows the fitted dispersions of the two inelastic features along ${\bf Q}=(h,0,22)=Q_a$ for \unit[0]{T} and \unit[2]{T}. The error bars reflect the uncertainty arising from the three-peak fitting procedure. The low-energy mode clearly disperses from the IC momentum value ${\bf Q}\!=\!(0.574,0,22)$ and reaches a maximum energy $\sim\unit[16]{meV}$ away from it. A slight softening can be seen near the point corresponding to the FIZZ state, ${\bf Q}\!=\!(0,0,22)$. In addition, the application of a \unit[2]{T} magnetic field softens the dispersion around that point. The higher broad continuum-like mode centered around $\sim\unit[35]{meV}$  is  almost non-dispersive and does not show any significant  changes in the applied field. It is noteworthy that recent independent studies of Ir L$_3$-edge RIXS on two related compounds, Na$_2$IrO$_3$ and $\alpha-$Li$_2$IrO$_3$, have observed a similar broad continuum around $\sim\unit[20-30]{meV}$, which has been interpreted as magnetic in origin through a careful study of its momentum and temperature dependence \cite{revelli_fingerprints_2020,kim_rixs_2020}.

The temperature dependence of the RIXS spectra is shown in Figure \ref{fig:fig3}\,c. The low-energy mode dispersing up to $\sim\unit[15]{meV}$ (indicated by the left arrow) monotonically decreases above $T_I$ and becomes featureless above $\sim\unit[150]{K}$. Conversely, the broad excitation mode centered around \unit[35]{meV} remains largely unchanged with increasing temperature. Its intensity, integrated for energies above \unit[20]{meV}, remains constant up to $\sim\unit[100]{K}$, and slowly decreases as $1/T$ above that. Nonetheless, this excitation persists to very high temperatures, with a sizable intensity at room temperature.

To better understand our data, we first model the ${\bf Q}$-dependence of the RIXS response using a semiclassical approach. To this end, we follow previous theoretical works which employ a $1/S$-expansion around the closest commensurate approximant of the observed IC order, namely a period-3 state with counter-rotating moments and the same irreducible representation as the observed IC order~\cite{ducatman_2018,rousochatzakis_magnetic_2018,Li_magneticfield_2020,li_reentrant_2020}. Expanding around this period-3 state leads to a magnon excitation spectrum and the spin DSF (\ref{DSF-main}).
Figure \ref{fig:fig3}\,b shows the diagonal components of the DSF $\mathcal{S}^{aa}({\bf Q},\omega)$, $\mathcal{S}^{bb}({\bf Q},\omega)$, $\mathcal{S}^{cc}({\bf Q},\omega)$, as well as the total DSF defined as  $\mathcal{S}({\bf Q},\omega)=\mathcal{S}^{aa}({\bf Q},\omega) + \mathcal{S}^{bb}({\bf Q},\omega)+\mathcal{S}^{cc}({\bf Q},\omega)$ along  the orthorhombic $Q_a$ direction, computed at $\mu_o$H$_b=0$T  (left panel) and $2$T (right  panel). Note that the off-diagonal components are non-zero, but they are several orders of magnitude weaker compared to the diagonal components, which is why we disregard them in our analysis. In the background we have also superimposed the linear spin wave spectrum (thin white lines). We can see that the most intense mode both at zero-field and at 2T is the one near ${\bf Q}=(2/3,0,0)$, which describes the fluctuations of the IC magnetic order. In particular, the dominant contribution to that mode comes from the $cc$ polarization channel. The intensities of the soft modes at ${\bf Q}\!=\!0$ are much weaker and are dominated by the $bb$ channel, consistent with the fact that the low-field ${\bf Q}\!=\!0$ zig-zag and ferromagnetic static orders are much weaker compared to the IC components. The applied field causes a softening of the ${\bf Q}\!=\!0$ mode compared to the zero-field data, as seen experimentally (more details in the Supplementary Information).

The magnetic continuum around $\sim\unit[35]{meV}$ is not seen in our semiclassical analysis of the DSF (see Figure \ref{fig:fig3}\,b). Moreover the fact that it survives at temperatures well above the ordering temperature $T_I$ with almost the same momentum-integrated intensity as the one at very low temperatures (see Figure \ref{fig:fig3}\,c) suggests that magnons cannot account for this broad continuum. The dominance of the Kitaev coupling in \blio{} offers an alternative interpretation, whereby the high-energy continuum of magnetic excitations is predominantly coming from long-lived pairs of Majorana fermions, characteristic of the proximate QSL, which is seen in the predominantly  SC response. In particular, according to theory calculations for the pure Kitaev model~\cite{Gabor2017}, the SC response of \blio{} is given by a broad continuum, centered at an energy around $6K/4\sim 27$ meV~\footnote{Note that we have rescaled the energy by a factor of four to account for the different convention of coupling constants in \cite{Gabor2017}.}, in good agreement with the observed position of the peak around 35\,meV. The difference is likely accounted for by the additional interactions in \blio{}, and in particular by $\Gamma$, which is as high as 55\% of the dominant Kitaev coupling.

\section{Discussion}

The low-energy excitation spectra hold important information about the nature of the magnetic state in \blio{}. With our data in agreement with semiclassical calculations of the dynamical spin structure factor, we can confidently use the model Hamiltonian to compute the expected excitation spectra at higher fields. 
In Figure \ref{fig:fig4}, we show the evolution of the low-energy spectrum at a wide range of magnetic field values H$_b$. We can see that the $Q_a=2/3$ mode remains soft and has significant intensity even at fields much larger than H$_b^*$, indicating strong spin fluctuations despite the absence of the corresponding long-range magnetic order. This happens because the modulated components with ${\bf Q}=(2/3,0,0)$ and the uniform components with ${\bf Q}=(0,0,0)$ are intertwined at all fields due to the spin-length constraints, as was previously shown by \citet{Li_field_2020}. With increasing field, this mode's energy increases continuously. However, only at fields above 100\,T does the spectrum show a dispersion with a minimum at  ${Q}=0$ wavevector characteristic of a fully polarized state. It is remarkable how the system resists full polarization due to the presence of intertwined orders.
  
On the other hand, the broad continuum around $\unit[35]{meV}$ shows clear clues on the impact of the dominant Kitaev interaction on the spectrum of \blio\ at intermediate and high energy scales, and reflects the near proximity of this system to the ideal quantum spin liquid state. The unusual temperature dependence of the observed broad continuum cannot be explained in terms of conventional magnon or phonons since the spin degree of freedom contribution to the scattering intensity should rapidly decrease and vanish at a temperature above the characteristic exchange energy, while a lattice contribution would increase monotonically with temperature due to the Bose population factor \cite{kim_rixs_2020}. However, the integrated intensity of this excitation shows Curie-like behavior ($1/T$) above \unit[100]{K} and a constant value below that, without being affected by the low-temperature long-range order. The thermal characteristics of the $JK\Gamma$ model have been theoretically studied and it is predicted that above the low-temperature order ($T>T_I$), itinerant Majorana fermions remain coherent up to a temperature $T\sim K$ where the fractional excitations recombine into spins and the nearest-neighbor spin correlations decay with increasing temperature. This fractionalization is experimentally observable in specific heat, thermal transport and spin correlation measurements, and its associated excitations are very robust against temperature. 

Recently, \citet{Ruiz_HighT_2020} reported a magnetic anomaly in \blio\, around $T_{\eta}=\unit[100]{K}$ with signatures appearing in magnetization, heat capacity, and muon spin relaxation ($\mu$SR) measurements. A sharp onset was observed on the magnetization at $T_{\eta}$ as well as a crossover in the heat capacity. Below \unit[100]{K}, magnetic hysteresis was detected which grows with decreasing temperature and penetrates the low-temperature magnetically ordered state with complete impunity. This temperature also coincides with previous report on the reordering of the principle magnetic axes in the three-dimensional iridates \cite{modic_lio_2014,ruiz_correlated_2017, majumder_nmr_2019} and the onset of fermionic contributions to the Raman spectra \cite{glamazda_raman_2016}. Moreover, $\mu$SR data indicate that the entity responsible for this transition-like anomaly is static and homogeneously distributed throughout the sample without causing a true long-range ordered state.

Our observation of the temperature dependence of the broad multi-spinon excitation centered at \unit[35]{meV} provides a connection to the previously reported magnetic anomaly at $T_{\eta}=\unit[100]{K}$ and suggests that $T_{\eta}$ may represent a crossover between a high-temperature featureless paramagnet and a proximate spin liquid regime governed by the physics of Majorana fermions. Below $T_{\eta}$, the nearest neighbor spin correlations saturate to their low-temperature value. These spin-spin correlations are precisely equivalent to the kinetic energy of the emergent Majorana fermions which release $1/2R\ln{2}$ of the spin entropy as the system enters this thermal crossover \cite{thermo_eschmann_2020}.
In addition, we note that similar RIXS excitations have been observed in the two-dimensional honeycomb iridates \anio\, and \alio\, \cite{revelli_fingerprints_2020, kim_rixs_2020}, which has been equated to the inelastic neutron scattering continuum reported on $\alpha-$RuCl$_3$. Such ubiquity across Kitaev materials is another confirmation of the dominant Kitaev exchange, and its characteristic multi-spinon-like dynamical fingerprints at intermediate energy and temperature scales, irrespective of the long range ordering taking place at very low temperatures.

\section{Conclusion}

We have studied the low-energy excitations of the three-dimensional Kitaev magnet \blio\, using a medium- and a high-resolution RIXS spectrometers with \unit[25]{meV} and \unit[10]{meV} resolution at the Ir $L_3$-edge, respectively. These measurements were carried out using \unit[0]{T} as well as a \unit[2]{T} applied magnetic field in order to access the intertwined IC and FIZZ state. In contrast to the case of hydrostatic pressure, an applied field does not disturb the relativistic $j_\mathrm{eff}=\nicefrac{1}{2}$ state. Moreover, the field-temperature dependence of the low-energy RIXS spectra reveal two distinct modes: (1) dispersive magnons branching from ${\bf Q}$-vectors corresponding to IC and FIZZ states and reaching a maximum energy $\sim\unit[16]{meV}$, comparable with the reported Kitaev exchange energy in \blio\, \cite{ducatman_2018,Li_field_2020}, and (2) a broad-continuum of magnetic excitations centered at \unit[35]{meV}, which is insensitive to the low-temperature order, remains constant up to \unit[100]{K} and slowly decreases above that. Indications for such a continuum response around \unit[30]{meV} have been previously reported by Raman experiments~\cite{glamazda_raman_2016}. Also, this continuum contribution was not seen in the THz measurements of Ref.~\cite{Majumder2020} because these went only up to \unit[80]{cm$^{-1}$}. 

 We have compared the experimental low-temperature dispersing magnons to the calculated spin dynamical structure factor above the closest period-3 approximant of the actual IC order and have shown that the dominant contribution to the $Q_a=2/3$ mode (corresponding to fluctuations of the IC order)  comes from the $cc$ polarization channel, whereas the $bb$ channel dominates the $Q=0$, FIZZ mode. Experimentally, we observe that a 2\,T magnetic field softens the dispersion around $Q=0$, in agreement with the calculation. The validity of the minimal microscopic Hamiltonian allows us to predict the evolution of the dynamical structural factor for large values of H$_b$. The persistent of the soft mode around $Q_a=2/3$ up to $\unit[100]{T}$ suggests that the IC and FIZZ states are highly entangled. Only at very high fields of $\unit[150]{T}$ do our calculations predict a quadratic dispersion indicative of a polarized state. 

Moreover, we showed that the overall intensity of the multi-spinon continuum around \unit[35]{meV} has an unusual temperature dependence with constant intensity below $T_{\eta}=\unit[100]{K}$, and a slow decrease above that. This characteristic temperature has been previously reported as the onset of a magnetic anomaly affecting thermodynamic variables without causing long-range magnetic order \cite{Ruiz_HighT_2020}. Both findings point towards unconventional magnetism and suggest that $T_{\eta}$ represents the thermal fractionalization of spins into disordered fluxes and Majorana fermion excitations which almost entirely dominate the thermodynamic response of \blio{}.


\section{Acknowledgements}

We thank Yi-Zhuang You, Tarun Grover, John McGreevy, Dan Arovas, Ken Burch, Yiping Wang and Gabor Halasz for fruitful discussions. Use of the Advanced Photon Source was supported by the U. S. Department of Energy, Office of Science, Office of Basic Energy Sciences, under Contract No. DE-AC02-06CH11357. Material synthesis and experimental measurements were supported by the Department of Energy, Office of Basic Energy Sciences, Materials Sciences and Engineering Division, under Contract No. DE-AC02-05CH11231. The work at LBNL is funded by the US Department of Energy, Office of Science, Office of Basic Energy Sciences, Materials Sciences and Engineering Division under Contract No. DE-AC02-05-CH11231 within the Quantum Materials Program (KC2202). Alejandro Ruiz acknowledges support from the University of California President's Postdoctoral Fellowship Program. Natalia B. Perkins and Mengqun Li were supported by the U.S. Department of Energy, Office of Science, Basic Energy Sciences under Award No. DE-SC0018056. 
N.P.B and I.Z. acknowledge the support of Harvey Mudd College. A.F. acknowledges support from the Alfred P. Sloan Fellowship in Physics.


%

\end{document}